\documentclass[useAMS,usenatbib]{mn2e}
\usepackage{graphicx,amsmath,multirow,amssymb}
\usepackage{subfig}
\usepackage{natbib}
\newcommand{\comment}[1]{}

\def\simgt{\lower.5ex\hbox{$\; \buildrel > \over \F sim \;$}}
\def\simlt{\lower.5ex\hbox{$\; \buildrel < \over \sim \;$}}

\title[Dual dust chemistry in LMC stars]{Do evolved stars in the LMC show dual dust chemistry?}

\author[E. Marini et al.]{
E. Marini$^{1,2}$\thanks{E-mail: ester.marini@uniroma3.it},
F. Dell'Agli$^{3,4}$, D. A. Garc\'{\i}a--Hern\'andez$^{3,4}$,
M. A. T. Groenewegen$^{5}$, 
\newauthor
S. Puccetti$^{6}$, P. Ventura$^{2}$, E. Villaver$^{7}$
\\
$^{1}$Dipartimento di Matematica e Fisica, Universit\'a degli Studi Roma Tre, via
della Vasca Navale 84, 00100, Roma\\
$^{2}$INAF, Osservatorio Astronomico di Roma, Via Frascati 33, 00077, Monte Porzio Catone, Italy\\
$^{3}$Instituto de Astrof\'{\i}sica de Canarias (IAC), E-38200 La Laguna, Tenerife, Spain \\
$^{4}$Departamento de Astrof\'{\i}sica, Universidad de La Laguna (ULL), E-38206 La Laguna, Tenerife, Spain \\
$^{5}$Koninklijke Sterrenwacht van Belgi{\"e}, Ringlaan 3, 1180 Brussels, Belgium\\
$^{6}$ASI, Via del Politecnico, 00133 Roma, Italy \\
$^{7}$Departamento de Fisica Teorica, Universidad Autonoma de Madrid, Cantoblanco 28049 Madrid, Spain\\
}

\begin{document}

\date{Accepted, Received; in original form }

\pagerange{\pageref{firstpage}--\pageref{lastpage}} \pubyear{2012}

\maketitle

\label{firstpage}

\begin{abstract}
We study a group of evolved M-stars in the Large Magellanic Cloud, characterized by 
a peculiar spectral energy distribution. While the $9.7~\mu$m feature arises from
 silicate particles, the whole infrared data 
seem to suggest the presence of an additional featureless dust species. We propose that 
the circumstellar envelopes of these sources 
are characterized by a dual dust chemistry, with an internal region, harbouring carbonaceous
particles, and an external zone, populated by silicate, iron and alumina dust
grains. Based on the comparison with results from stellar modelling that describe the dust 
formation process, we deduce that these stars descend from low-mass ($M < 2~M_{\odot}$) 
objects, formed $1-4$ Gyr ago, currently evolving either in the post-AGB phase or 
through an after-pulse phase, when
the shell CNO nuclear activity is temporarily extinguished. Possible observations able
to confirm or disregard the present hypothesis are discussed.
\end{abstract}




\section{Introduction}
The Large Magellanic Cloud (LMC) has been so far the most
investigated environment to test asymptotic giant branch (AGB) evolution modelling, 
owing to its relative proximity \citep[$\sim 50$ kpc,][]{feast99} 
and low average reddening \citep[$E(B-V) \sim 0.075$,][]{schlegel98}.

\begin{figure*}
\begin{minipage}{0.30\textwidth}
\resizebox{1.\hsize}{!}{\includegraphics{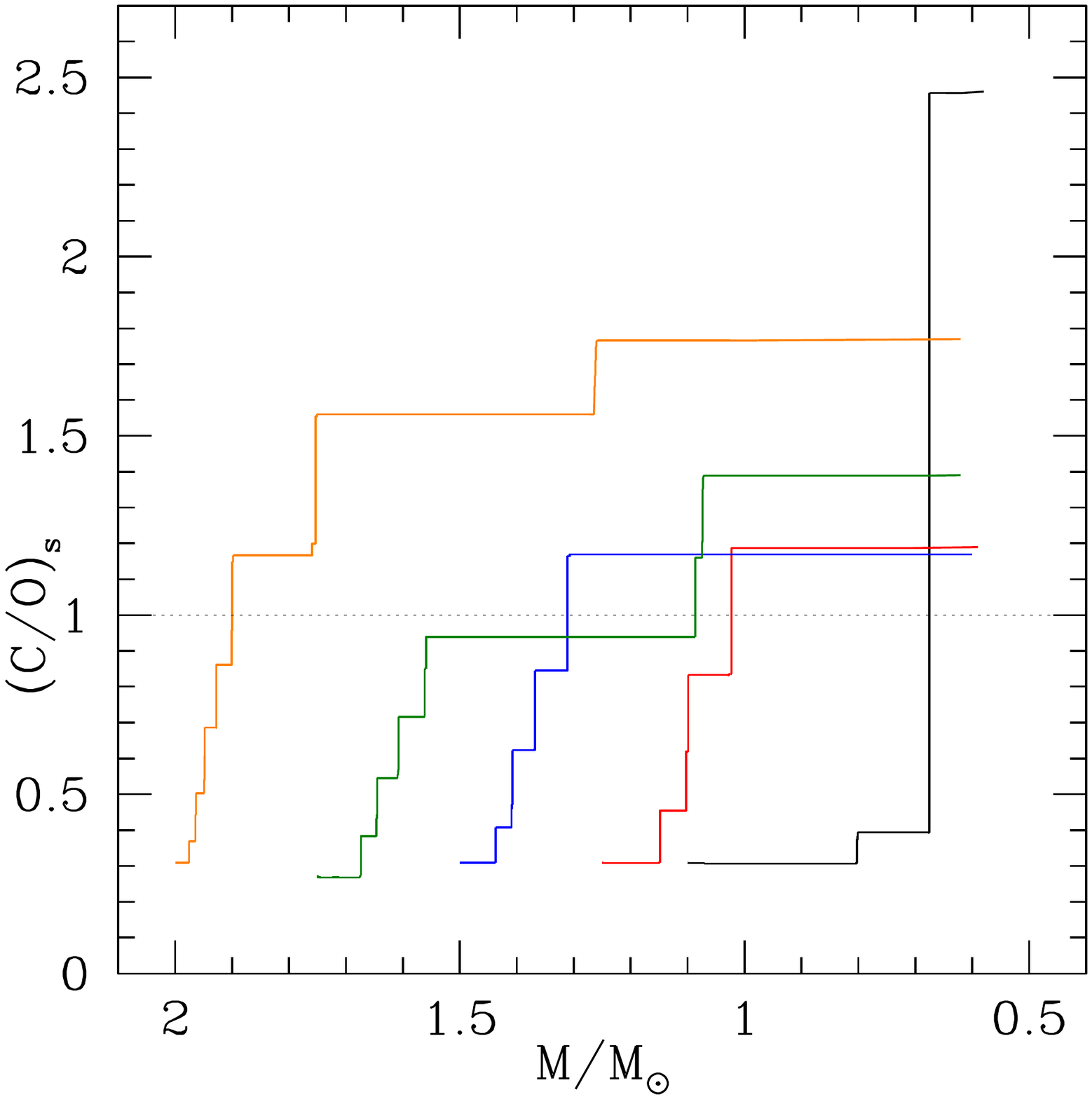}}
\end{minipage}
\begin{minipage}{0.30\textwidth}
\resizebox{1.\hsize}{!}{\includegraphics{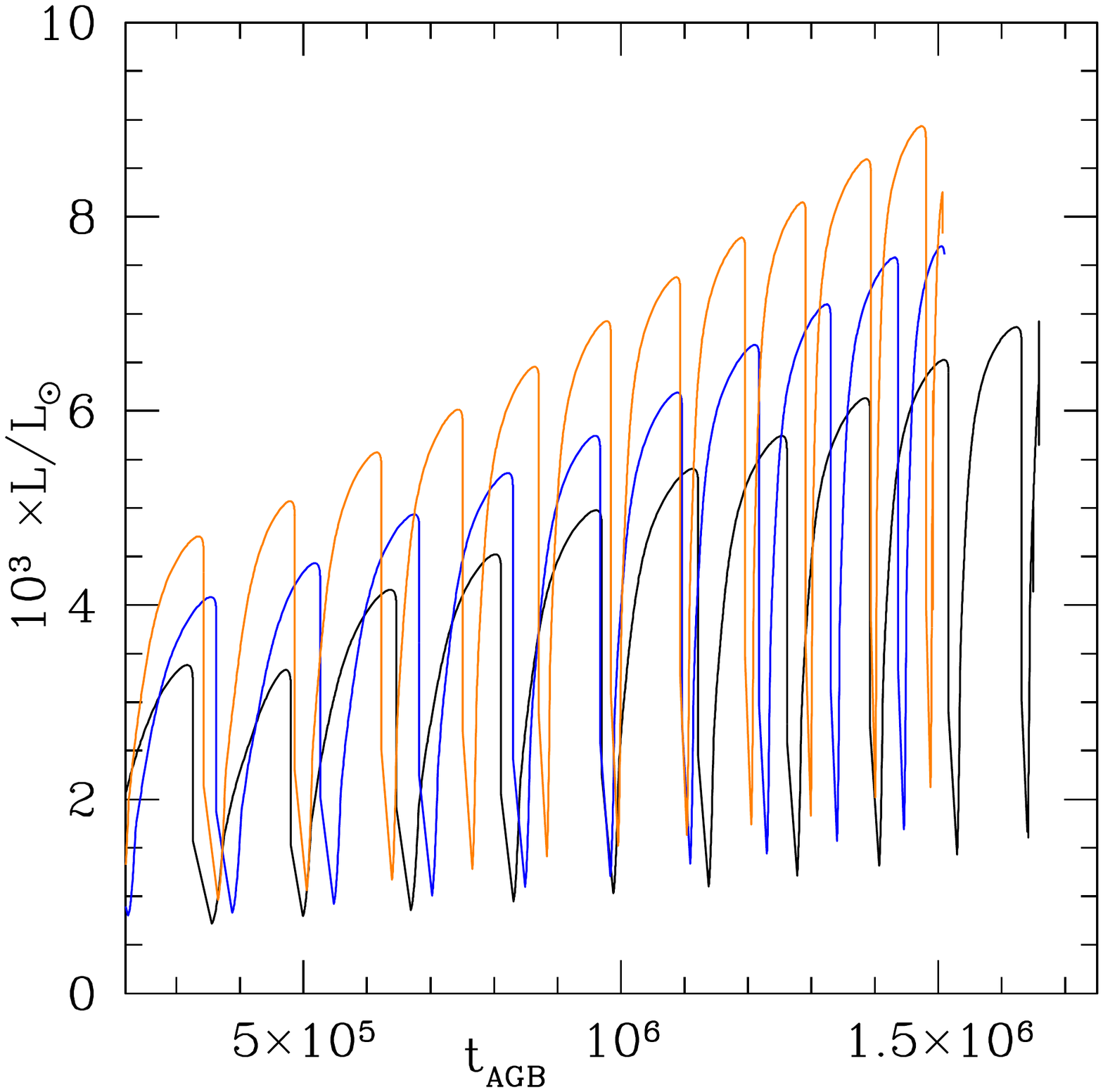}}
\end{minipage}
\begin{minipage}{0.30\textwidth}
\resizebox{1.\hsize}{!}{\includegraphics{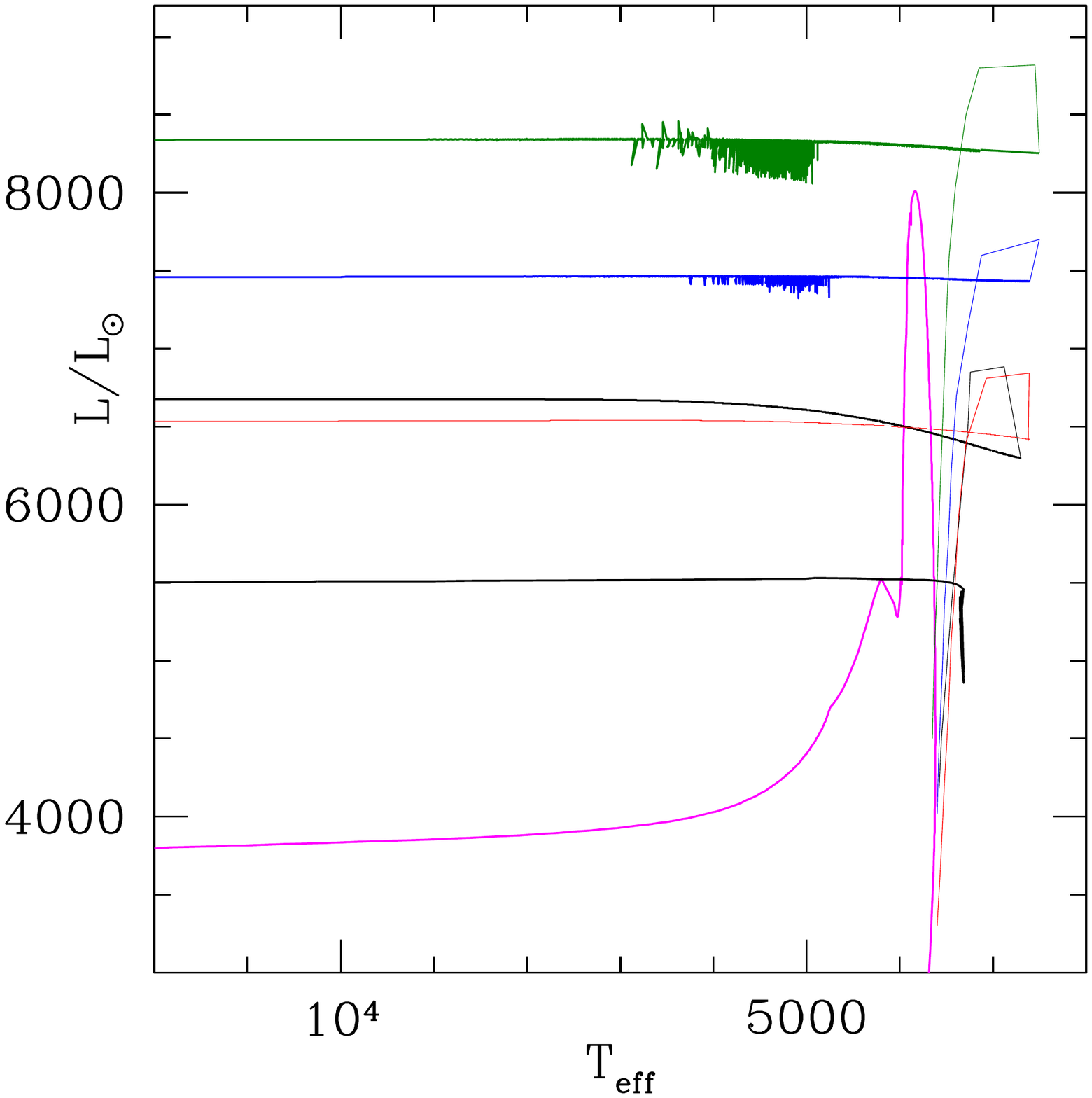}}
\end{minipage}
\vskip-40pt
\caption{Left: the variation of the surface C$/$O ratio during the AGB evolution of stars
of initial masses $1.2~M_{\odot}$ (black line), $1.3~M_{\odot}$ (red), $1.5~M_{\odot}$ (blue),
$1.75~M_{\odot}$ (green), $2~M_{\odot}$ (orange). Middle: the evolution of the luminosity
of the $1.2~M_{\odot}$, $1.5~M_{\odot}$ and $2~M_{\odot}$ stars. Right: The post-AGB evolution
of the stars of mass below $2~M_{\odot}$ in the HR diagram. The pink line corresponds to a star
of mass $1~M_{\odot}$ and the light lines on the right 
refer to the AGB phase and connect the temperatures and luminosities of the stars 
during the interpulses.}
\label{fmod}
\end{figure*}

The Surveying the Agents of a Galaxy's Evolution Survey (SAGE,
Meixner et al. 2006), with the \emph{Spitzer Space Telescope}, 
provided mid-IR data of millions of AGB stars, that allowed to 
study the dust enrichment from the individual sources \citep{srinivasan09} and to
estimate the current dust production rate by AGB stars 
in the LMC \citep{mikako,raffa14}.

Among the results from the \emph{Spitzer} space mission we stress the importance
of the SAGE-Spec atlas \citep{kemper10}, obtained with the Infrared Spectrograph (IRS).
The availability of the spectral energy distribution (SED) in the whole mid-IR
region offers a valuable opportunity of determining the luminosity and the dust 
composition within the circumstellar envelope of the individual sources 
observed. The latest developments in AGB modelling, with the evolution of the central 
star coupled to the description of dust formation in the wind, allow 
the characterization of the individual sources, in terms of the properties 
of the progenitors and the evolutionary phase \citep{flavia15a, flavia15b}.

This research will be of paramount importance for the incoming James Webb Space 
Telescope (JWST) mission, because the analysis so far being limited to the Magellanic 
Clouds (MCs) can be extended to all the galaxies in the Local Group. The recent studies 
by \citet{jones14,jones17} are extremely important in this regard, as they provided 
the magnitudes of obscured M-stars in the LMC that would be found if they were observed 
with the JWST filters.

Against this background we decided to extend the analysis by \citet{jones14}, using
the results from AGB modelling to interpret the IRS SED and to characterize
the individual sources. In a first study we focused on a few
stars with a peculiar SED, which we interpreted as massive metal-poor AGB stars, 
whose dust composition is dominated by solid iron particles \citep{ester19}.

In this paper we focus on 8 sources, which we propose to be characterized by a dual dust 
chemistry, surrounded by an internal dust layer and an external shell, populated, 
respectively, by carbonaceous particles and silicate grains. 

The idea that AGB stars with dual-chemistry exist was proposed independently by 
\citet{willems86} and \citet{little86}, after the detection of objects that showed clear 
absorption bands from carbon-rich molecules in their optical and near-IR spectra while 
showing silicate emission in the mid-IR. The possible presence of such systems is challenged by the expansion of the
outer silicate-rich layer, to the point where the silicate feature is no
longer detectable \citep{chan88}. On the other hand, the formation of a disc, possibly 
related to the presence of a companion, might prevent the escape of the
external silicate shell \citep{lloyd90}.
Here we try to identify the physical conditions and the evolutionary phases allowing
the formation of this class of objects.

These findings will be also important to 
identify dual dust chemistry stars in samples of evolved stars in galaxies.

\section{Dust formation in low-mass AGB stars}
We used the AGB models discussed by
\citet{ventura14}, where the description of dust formation in the wind 
is coupled to the evolution of the central stars, i.e. the variation of the main
physical properties (luminosity, effective temperature, mass loss rate) and the
surface chemical composition. In the present work we use models with
$Z=8\times 10^{-3}$, the dominant metallicity in the LMC \citep{harris09} that were presented 
by \citet{flavia14, flavia15a, flavia15b}. For the scope of the 
present investigation, we extended the tracks of the stars of initial mass below 
$2~M_{\odot}$ to the post-AGB phase, up to effective temperatures $\sim 5\times 10^4$ K
(these results will be presented in a forthcoming paper).

Fig.~\ref{fmod} shows the AGB evolution of stars of initial mass 
$1.2~M_{\odot} \leq M \leq 2~M_{\odot}$, all reaching the C-star stage. We 
report the variation of the surface C$/$O ratio (left panel), the luminosity (middle) and 
the excursion of the tracks in the HR diagram during the post-AGB phase (right). 

The stars of mass below $2~M_{\odot}$ experience a last thermal pulse (TP), after which they
become C-stars, then start the post-AGB evolution, without experiencing further TPs.
This is because after the C$/$O$>1$ condition is reached a significant increase in the 
rate of mass loss speeds up the loss of the envelope \citep{eric08}, thus preventing 
additional TPs.

These stars produces oxygen-rich dust for most of their lives on the AGB, but the phases 
following the last TP, when they produce carbonaceous dust. 
The change in the dust mineralogy associated with the occurrence of the TP 
set the conditions for the presence of two dusty shells, with amorphous carbon and
silicate dust.
These results indicate that the hypothesis that a late TP converts an M-star into a 
C-star, proposed by \citet{perea09} to explain the dual 
chemistry phenomenon in Galactic bulge planetary nebulae (hereafter PNe), is indeed a common behaviour of all the stars 
formed between 1 and 4 Gyr ago.\\
\begin{figure*}
\begin{minipage}{0.33\textwidth}
\resizebox{1.\hsize}{!}{\includegraphics{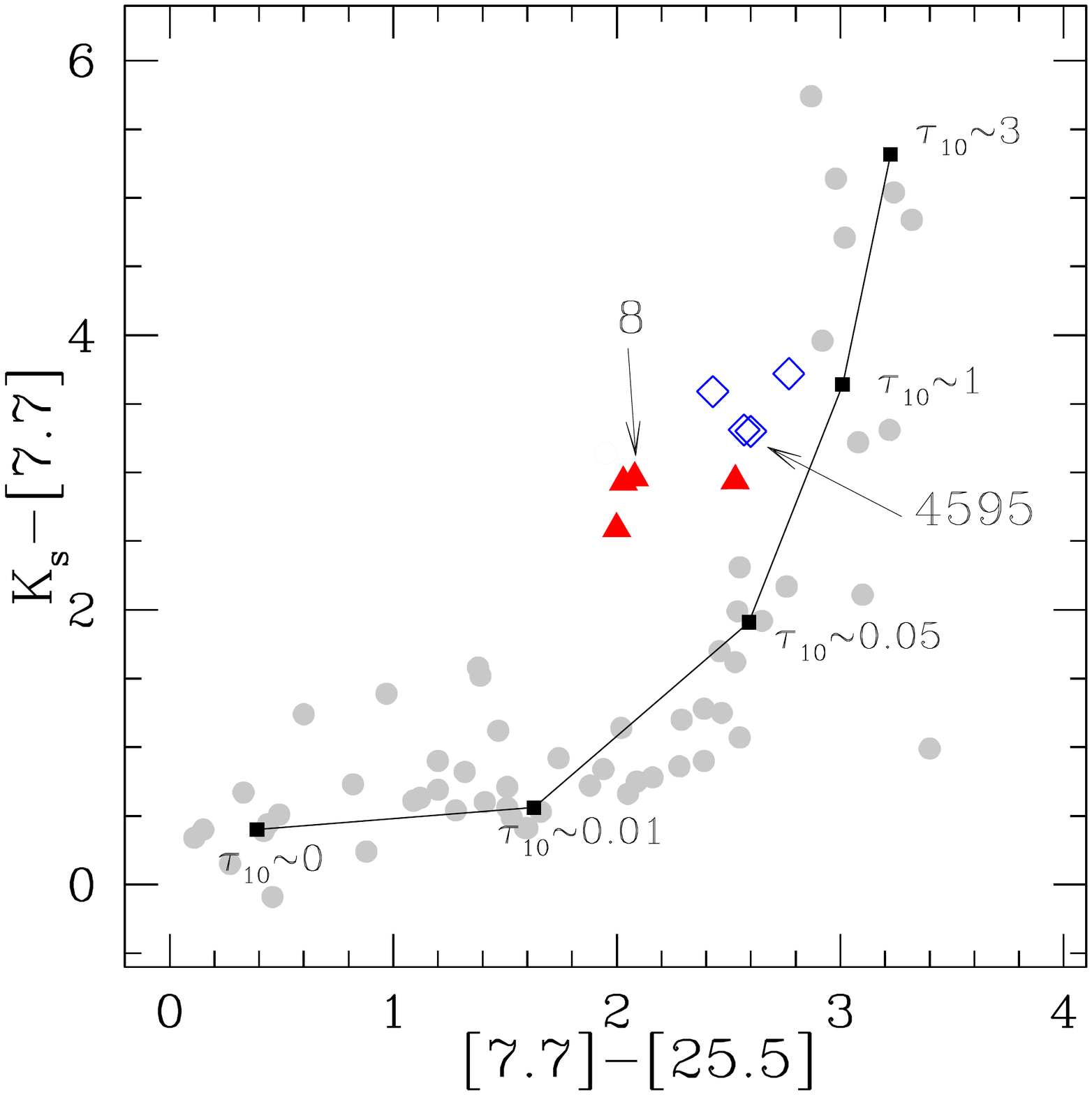}}
\end{minipage}
\begin{minipage}{0.33\textwidth}
\resizebox{1.\hsize}{!}{\includegraphics{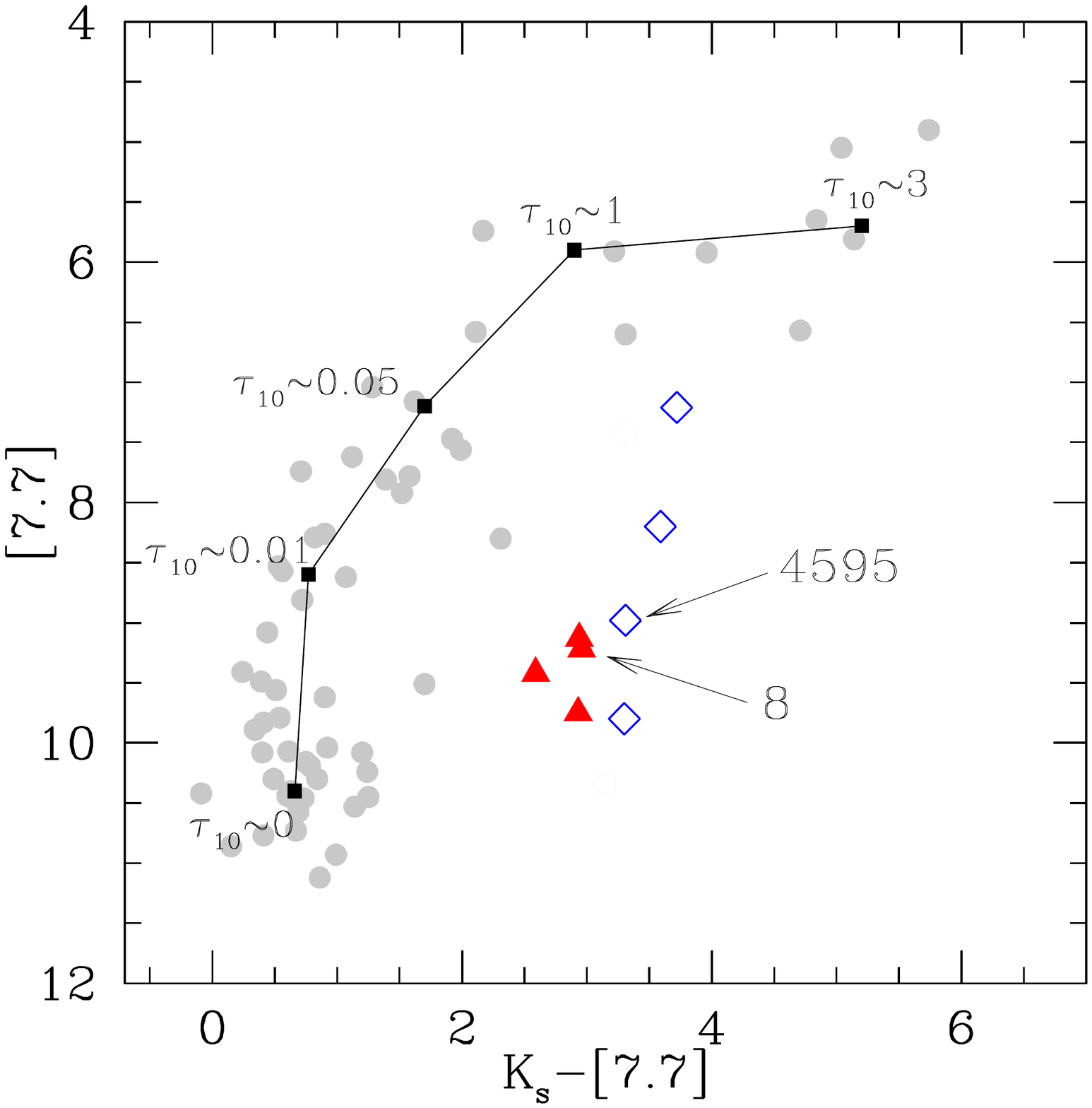}}
\end{minipage}
\begin{minipage}{0.33\textwidth}
\resizebox{1.\hsize}{!}{\includegraphics{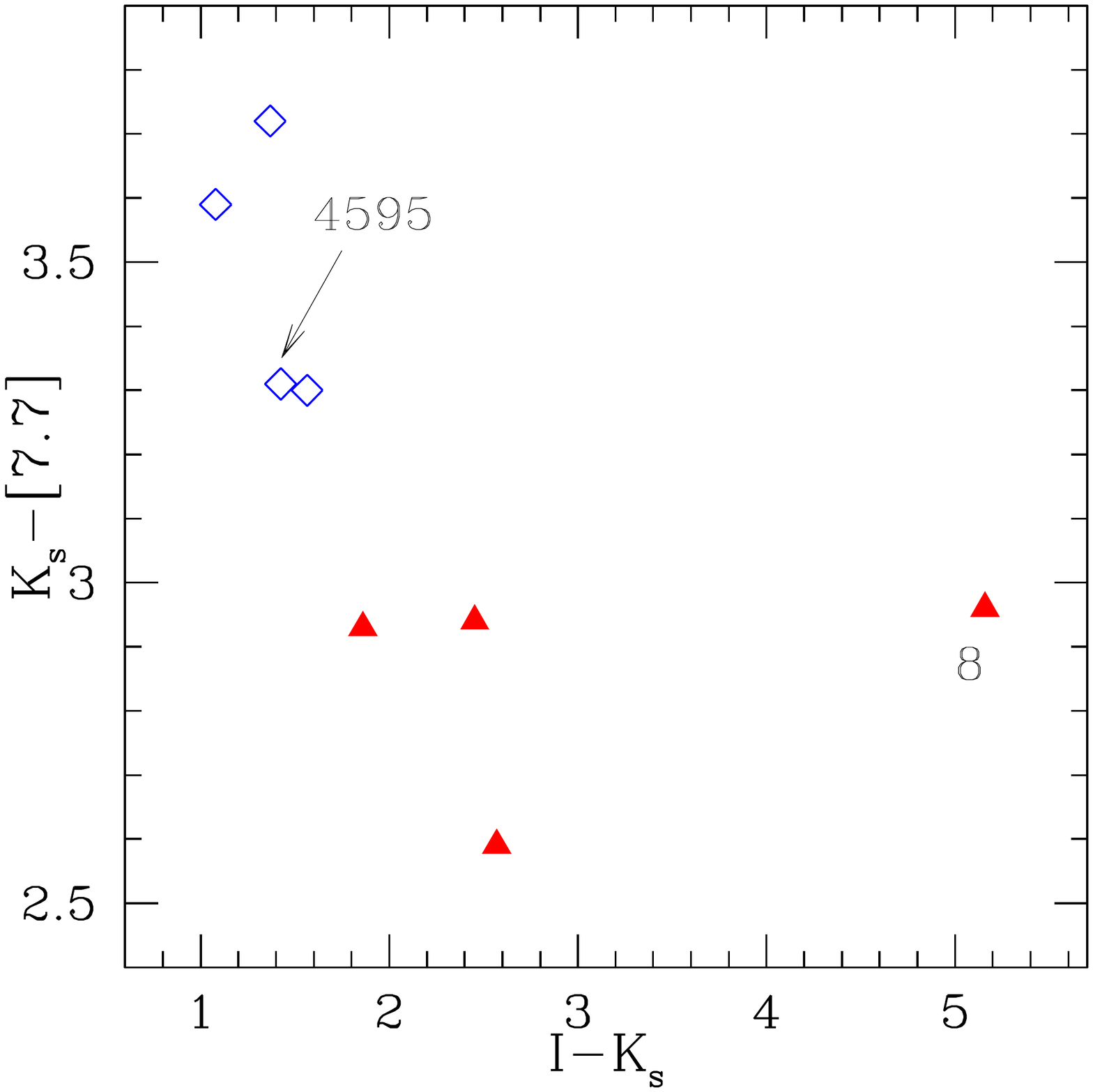}}
\end{minipage}
\vskip-40pt
\caption{Distribution of the sample of M-stars in the LMC analyzed by \citet{jones17}, 
indicated by grey points, in the colour-colour (left panel) 
and in the colour-magnitude diagrams (middle panel). Blue open diamonds and 
red full triangles indicate the two groups of stars discussed in the present paper.
The black lines indicate a theoretical obscuration sequence of M-stars, assuming 
a dust composition made up of silicates, alumina dust and solid iron. Typical values
of the optical depth at $10~\mu$m and the position of the sources
SSID 8 and SSID 4595 are indicated. In the right panel is shown a diagram with a colours combination able to clearly separate the stars among the two groups described in sections 3.1 and 3.2. $I$ magnitudes come from \citet{zarit04}.
} 
\label{fcol}
\end{figure*}
M$\geq 2~M_{\odot}$ behave differently,
because they accumulate large quantities of carbon in the photosphere,
that favour the formation of great quantities of carbonaceous dust \citep{eric08, flavia15a}.\\
Fig.~\ref{fmod} shows that the luminosities experienced during
the inter-pulse phase of the AGB life are in the range $3-10 \times 10^3~L_{\odot}$, while
the C-star stage is reached when $5\times 10^3~L_{\odot} < L < 10^4~L_{\odot}$. 
During the phases following the ignition of the TP, when the CNO nuclear 
activity in the shell is temporary extinguished, the luminosities drop to 
$\sim 2\times 10^3~L_{\odot}$.
The right panel of Fig.~\ref{fmod} shows that the luminosities attained
during the post-AGB phase are in the range $6\times 10^3~L_{\odot} < L < 9\times 10^3~L_{\odot}$,
and stay approximately constant as the evolutionary tracks move to hotter effective
temperatures. These results are consistent with the $Z=0.01$ post-AGB
models published by \citet{marcelo16}.

\section{Dual dust chemistry M-stars}
We consider the M stars in the LMC with IRS spectra available, selected by 
\citet{jones14}. To study the distribution of the
observational planes we use the mid-IR magnitudes calculated by \citet{jones17}, 
who integrated the SED of each source over the MIRI spectral response. 

The left panel of Fig.~\ref{fcol} shows the distribution of these stars in the 
colour-colour plane ($[7.7]-[25.5]$, $K_S-[7.7]$) (we expect to obtain similar 
trends when using the NIRCam F210M filter). The stars trace an obscuration sequence,
up to colours $[7.7]-[25.5] \sim 3$ and $K_S-[7.7] \sim 6$. An obscuration path is 
also present in the colour-magnitude diagram ($K_S-[7.7]$, $[7.7]$) (see middle panel of
Fig.~\ref{fcol}). \\
The stars highlighted with colour symbols fall off
the obscuration sequences. This is particularly evident in the middle panel, 
where these sources define a vertical sequence below the main obscuration 
pattern, at $K_S-[7.7] \sim 2.5-3.5$.\\
The peculiar position of these objects in the observational planes is due to their
SED: while the $9.7~\mu$m feature indicates the presence of silicates, the continuum in the 
$\lambda < 8~\mu$m region suggests the presence of a featureless dust species. The two most
plausible options are solid iron and amorphous carbon, given the low stability and the
small abundances of other species. A dominant contribution from solid
iron to the overall extinction was explored by \citet{ester19} in the context of low-metallicity,
massive AGB stars, in which strong hot bottom burning (hereafter HBB) inhibits the formation of silicates, via the
destruction of the surface oxygen and magnesium. This possibility can be ruled out in the
present case, because the luminosities of the stars analyzed here, below $10^4$ L$_{\odot}$, are
much smaller than those typical of the stars undergoing HBB, above $2\times 10^4$ L$_{\odot}$. 
Therefore, we will base our analysis on the hypothesis of the presence of a dust layer
composed by amorphous carbon.\\
In the following we divide these stars into two groups, according to 
the morphology of their SED, derived from IRS and photometric data. 
This classification can be appreciated in the colour-colour diagram 
($I-K_S$, $K_S-[7.7]$) shown in the right panel of Fig.~\ref{fcol}, in which these two 
groups clearly separate from each other.

\begin{figure*}
\begin{minipage}{0.44\textwidth}
\resizebox{1.\hsize}{!}{\includegraphics{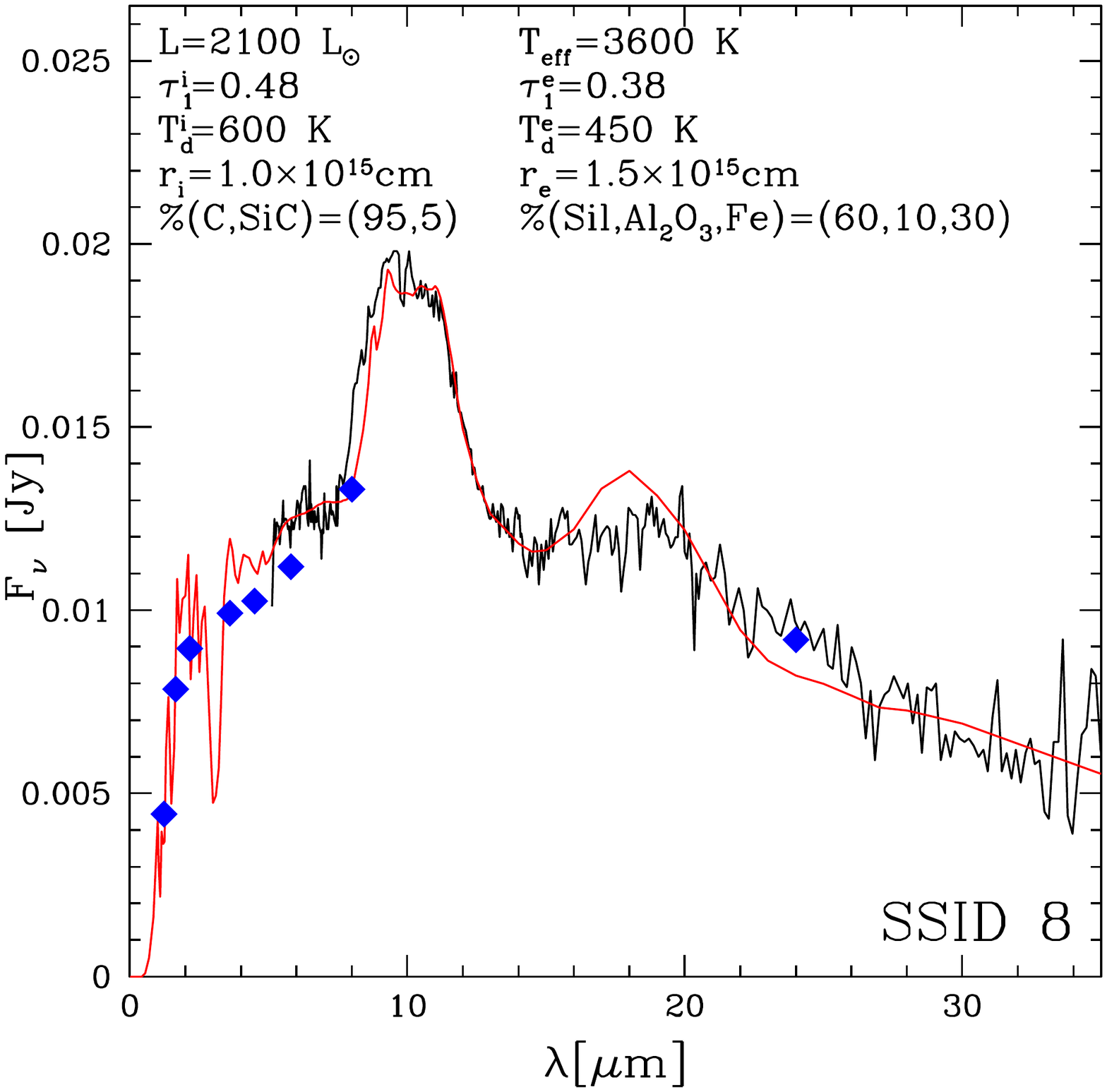}}
\end{minipage}
\begin{minipage}{0.44\textwidth}
\resizebox{1.\hsize}{!}{\includegraphics{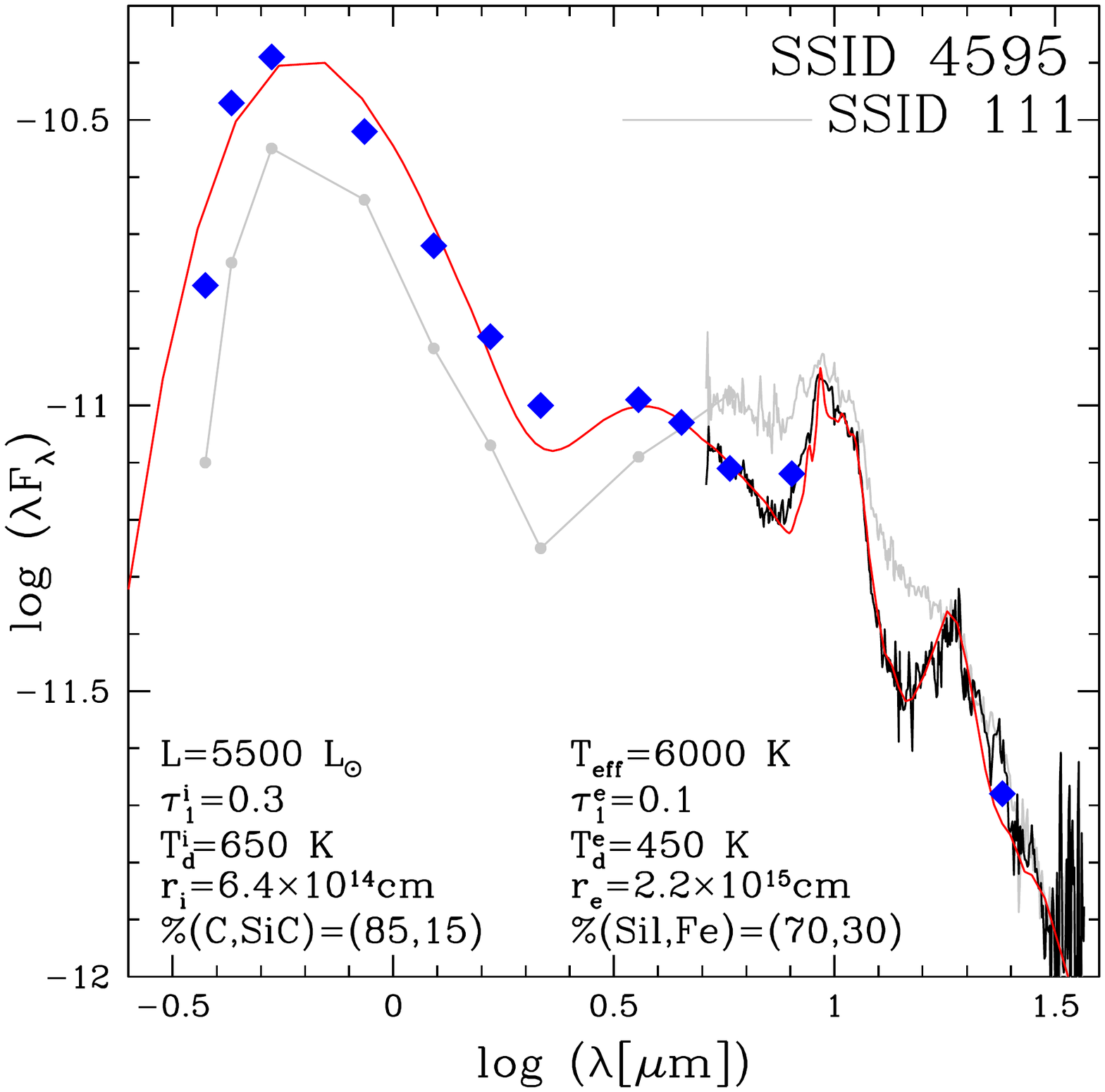}}
\end{minipage}
\vskip-58pt
\caption{The comparison between the observed IRS spectra (in black) and the synthetic 
SED (in red) of the stars in the LMC sample by \citet{woods11}: SSID 8 (left panel, 
see Fig.~\ref{fcol}) and SSID 4595 (right). Blue diamonds show \emph{Spitzer} photometry
from \citet{meixner06} and optical + near IR data, available in the literature. 
The grey line in the right panel refers to SSID 111, classified as a post-AGB stars
from \citet{woods11}. In both panels we report the luminosity and effective
temperature of the star, and the following parameters for the internal and outer
dust shells: optical depth (at $1~\mu$m), dust temperature, 
dust composition and distance from the central star. For SSID 8 we modelled the 
emission from the central star with a GRAMS atmosphere corresponding to the best-fit 
effective temperature, while for SSID 4595 we used a black body spectrum.}
\label{fsed}
\end{figure*}

\subsection{A class of post-AGB stars}
The sources plotted with blue open diamonds in Fig.~\ref{fcol} exhibit a short-wavelength 
peak, at $\sim 0.4-0.6~\mu$m, and two features, at 
$9.7~\mu$m and $18~\mu$m, that reveal the presence of silicates. An example
is shown in the right panel of Fig.~\ref{fsed}.
This SED indicates a post-AGB nature, because the effective
temperatures deduce from SED fitting are above
$\sim 5000$ K. \\
We propose that these post-AGB stars descend from $1.2 \leq M/M_{\odot} < 2$ 
progenitors. A nice fit of the observed SED (see right panel of Fig.~\ref{fsed})
is obtained by assuming the presence of two dusty layers: the more internal is composed 
by carbon dust, whereas the more external hosts mainly silicate particles. 
A similar explanation was invoked by \citet{bunzel09}
to interpret the SEDs of Galactic O-rich post-AGB stars (heavily obscured and
showing strong silicate absorption) although their peculiar SEDs could be
fitted with pure silicate dust when including very large and cold grains (R.
Szczerba 2019, priv. comm.); a possibility that we rule out for the significantly less
obscured stars discussed here.

Based on the discussion in the previous section, we believe that the two dust layers 
formed just before and after the last TP, when the transition from M-star to C-star 
occurred. 
This interpretation is supported by the luminosities required to reproduce 
the SED, in the range $5-10 \times 10^3~L_{\odot}$, in agreement with the post-AGB 
luminosities reported in the right panel of Fig.~\ref{fmod}. 
This understanding  is in agreement with the fact that only one source shows a 90-day
period, and the other 7 show no signs of variability \citep{jones17}. A further probe for this interpretation comes from the presence of a weak PAH-like emission feature at $6.3~\mu m$ in a post-AGB star in our sample
(SSID 4547), confirming the
presence of C-rich material and the classification as post-AGB stars\footnote{A few PNe with a similar dual-dust chemistry (i.e., PAHs + amorphous silicates) have been observed in our Galaxy (e.g. Perea-Calder{\'o}n et al. 2009). Curiously, their $6.3~\mu m$
feature is always the strongest one, as seen in our star SSID 4547.}.\\
The presence of these stars in the AGB sample is due to the criterion
followed to distinguish AGB from post-AGB stars, discussed by \citet{woods11} (see
their Fig.~3). Post-AGB stars are commonly identified by the presence of two 
distinct peaks in the SED, located in the $\lambda < 1~\mu$m portion of the spectrum 
and at $9.7~\mu$m, separated by a clear minimum in the SED at $\sim 2~\mu$m. The presence 
of carbon dust affects significantly the shape of the SED in the $1-8~\mu$m region, 
preventing the appearance of a minimum, making the SED very similar to a M-type AGB. 

The Milky Way is known to host dual chemistry PNe \citep{Gutenkunst08, lizette11, gg2014} 
and post-AGB stars \citep{cerrigone09, waelkens86, molster01}; however, no dual chemistry 
PNe or post-AGB stars have been previously detected in the MCs \citep{letizia2007, bernard09}, 
which could pose a problem to the present suggestion of the existence of dual chemistry 
post-AGBs. However, this might be just a consequence of small-number statistics given the 
low number of PNe that has been observed in the MCs in the infrared compared with the 
number of AGBs stars.  

\subsection{Faint, dusty AGB stars}
The SED of the stars indicated with red full triangles in Fig.~\ref{fcol} show 
evidence of cool dust, with a peak at $9.7~\mu$m,
revealing the presence of silicate grains. 
The left panel of Fig.~\ref{fsed} shows an example of these spectra. The 
luminosities deduced from SED fitting fall in the range
$2-4\times 10^3~L_{\odot}$. 

Obscured M stars are produced either from $M > 3~M_{\odot}$ stars experiencing HBB or to low mass
stars (initial masses below $\sim 1.5~M_{\odot}$) evolving through advanced AGB phases, 
before becoming carbon stars. Both explanations are not plausible in this case, because 
the luminosities would be much in excess of those observed. Furthermore,  
we have not found a way of reproducing the SED in the $4-8~\mu$m 
region, without assuming significant amounts of carbon dust. \\
As discussed in the previous section (see middle panel of Fig.~\ref{fmod}),
the luminosities given above are compatible with the post-TP phases of low-mass
stars. During these phases dust formation stops, because the drop 
in the mass loss rate provokes a significant decrease in the density of gaseous
particles available to condensation.\\
We propose that these stars have reached the C-star stage via a TDU episode that followed 
the previous TP, and are currently evolving through a post-TP, low-luminosity
phase. Similarly to the post-AGB stars discussed earlier in this section, we reproduce
their SED by two dusty layers, with a more internal zone, 
populated by carbonaceous particles, and an outer region, with silicate grains. As shown in 
the left panel of Fig.~\ref{fsed}, the synthetic SED obtained with these assumptions 
reproduces the IRS and photometric data well. Both dusty layers
are cool, in agreement with the expectation that dust formation is halted during these
low-luminosity phases. The source shown in the left panel of Fig.~\ref{fsed} was
studied by \citet{martin18}, who tried to fit its SED by adopting a single silicate
dust shell. As shown in the two bottom left panels on page 77 of that paper, some
significant differences between the observed SED and the best fit exist. 
The dual-chemistry model, compared to the single silicate dust shell, agrees better, 
particularly in the location of the primary peak and in the flux in the
$\lambda > 15~\mu$m domain.

\section{Discussion}
According to our interpretation the two groups of stars discussed here have a common 
origin, as all of them descend from progenitors of mass $1~M_{\odot} < M < 2~M_{\odot}$,
consistent with the episode of star formation that the LMC experienced $3-5$ Gyr ago 
(Bertelli et al. 1992). They are characterized by a dual dust chemistry, being
surrounded by a dust layer composed of carbonaceous particles, and a more external zone,
harbouring silicates, alumina dust and solid iron. 
  
The presence of amorphous carbon, a featureless species, renders the SED of these objects
significantly different from stars of similar obscuration and luminosity
surrounded by O-bearing dust species only. 

We believe that the dual dust chemistry originates from the transition to a C-rich 
photosphere during the final phase of life on the AGB. If this understanding is correct, 
we would expect these stars are C-rich, with a surface C$/$O not far in excess of unity. 
Considering the general low signal-to-noise of  
the IRS spectra, there are no apparently real or strong C$_2$H$_2$ 
absorptions  at $7.5~\mu$m and $13.7~\mu$m in the stars proposed here; however, this is not surprising, as the optical depths of the
carbon dust layers derived here, of the order of $\tau_1 \sim 0.3-0.7$, make these 
features not clearly identifiable in the SED. Optical and/or NIR high-resolution 
spectroscopy would be 
valuable to confirm or disregard the present interpretation.

If confirmed, the results presented in this paper suggest that evolved stars with dual chemistry are expected
to be found in galaxies where significant star formation in the epochs $1-4$ Gyr ago
occurred. Their identification will be possible in the context of the JWST mission,
considering that they occupy specific regions in the various observational planes.

\section*{Acknowledgements}
FDA and DAGH acknowledge support from the State Research Agency (AEI)  
of the  Spanish Ministry of Science, Innovation and Universities  
(MCIU) and the European Regional Development Fund (FEDER) under grant  
AYA2017-88254-P.








\bsp	
\end{document}